# Long-Lived States of Methylene Protons in Achiral Molecules

Anna Sonnefeld[a], Aiky Razanahoera[a], Geoffrey Bodenhausen[a], Kirill Sheberstov*[a]

Dedication: To the memory of Ray Freeman

[a]  A. Sonnefeld, A. Razanahoera, Prof. G. Bodenhausen, Dr. K. Sheberstov
Department of chemistry
École normale supérieure, PSL University,
24 Rue Lhomond, 75005 Paris (France)
E-mail: kirill.sheberstov@ens.psl.eu

**Abstract:** It is shown that long-lived states (LLS) involving pairs of methylene protons in nuclear magnetic resonance (NMR) can be readily excited in common molecules that contain two or more neighboring $CH_2$ groups. Accessing such LLS does not require any isotopic enrichment, nor does it require any stereogenic centers to lift the chemical equivalence of $CH_2$ protons. The excitation of LLS is achieved by polychromatic spinlock induced crossing (poly-SLIC). In a variety of metabolites, neurotransmitters, vitamins, amino acids, and other molecules, LLS were created for the first time. One can excite LLS in several different molecules simultaneously. The lifetimes $T_{LLS}$ are typically 3 to 6 times longer than $T_1$.

## Introduction

In nuclear magnetic resonance (NMR), the memory of spin systems is normally limited by longitudinal relaxation. However, long-lived states (LLS) have lifetimes $T_{LLS}$ that can be significantly longer than $T_1$ [1–3]. The invention of long-lived states has opened new perspectives, in particular for revealing interactions between potential drugs and target proteins[4–6], for the observation of hyperpolarized metabolites[7–10], for probing slow chemical exchange[11], for determining rates of slow diffusion[12,13], for storing nuclear hyperpolarization[14–17], for selecting signals of interest in peptides and proteins[18,19], and for detecting signals of metabolites in magnetic resonance imaging (MRI) [20–22].

In molecules with two magnetically equivalent spins, like in gaseous hydrogen $H_2$ or in $CH_2Cl_2$, the two proton spins can be predominantly in the singlet state $\alpha\beta-\beta\alpha$, thus giving rise to a singlet-triplet population imbalance $\Delta P_{STI}$. This can persist for a long lifetime $T_{LLS}$ because the conversion of populations between the singlet and triplet manifolds is forbidden by symmetry. In molecules with $CH_2$ groups, a population imbalance $\Delta P_{STI}$ between the singlet and triplet manifolds of the two protons can be readily excited when the chemical equivalence of the two nuclei is lifted, i.e., when they have distinct chemical shifts[23]. This occurs for diastereotopic pairs of protons in chiral molecules, where the degeneracy of the chemical shifts is lifted by the presence of (possibly remote) stereogenic centers[24,25]. However, when the chemical shifts differ, the lifetime $T_{LLS}$ will be shortened by singlet-triplet leakage. This effect can be attenuated by shuttling the sample to low fields, or by 'sustaining' the LLS by strong radio-frequency irradiation at high fields. Both approaches suffer from obvious disadvantages.

This paper shows that one can also create LLS involving two methylene protons that are *chemically equivalent* (i.e., have identical chemical shifts), provided they are *magnetically* inequivalent, i.e., have distinct scalar couplings to other nuclei, such as the protons of nearby $CH_2$ groups. Pairs of geminal protons in $CH_2$ groups are good candidates for supporting population imbalances with long lifetimes, since the intrapair dipole-dipole interaction does not cause any relaxation of their population imbalance. Singlet-triplet imbalances are "protected" against coherent dissipation by geminal *J*-couplings[26] that lifts the degeneracy between the singlet and the central triplet states. The LLS of methylene protons can be excited by spin-lock induced crossing [25,27–29] as well as it's polychromatic version (poly-SLIC)[30]. These states are delocalized, since radio-frequency irradiation at the chemical shift of a selected $CH_2$ group also excites LLS associated with one to two neighbouring $CH_2$ groups. This unusual feature is due to the delocalized nature of the eigenstates [30].

## Results and Discussion

A few common molecules with aliphatic chains where LLS can be excited by SLIC are shown in **Figure 1**. The parameters optimized for LLS excitation and some relaxation properties are summarised in **Table 1**. The conditions to excite LLS are similar for all compounds, so that many independent LLS can be excited simultaneously in samples containing mixtures of molecules with aliphatic chains.

### Excitation of singlet states in methylene chains

Magnetic inequivalence between the protons of a methylene group is due to differences in the out-of-pair *J*-couplings. A spin system with two magnetically inequivalent $CH_2$ groups can be classified as $AA'XX'$ in Pople's notation [31]. Except if all rotamer populations are equal, the vicinal *J*-couplings are not degenerate, i.e., $^3J_{AX} = {}^3J_{A'X'} \neq {}^3J_{AX'} = {}^3J_{A'X}$. This makes it possible to populate a LLS of the strongly coupled system using SLIC. The optimum radio-frequency (RF) amplitude (nutation frequency) $\nu_1^{SLIC}$ and duration of the SLIC pulse depend on only two parameters:

$$2J_{intra} = {}^2J_{AA'} + {}^2J_{XX'}$$
$$\Delta J = |{}^3J_{AX} - {}^3J_{AX'}|$$
Eq. 1

The RF amplitude should match the sum of intrapair *J*-couplings $2J_{intra}$, while the duration $\tau_{SLIC}$ should be inversely proportional to the difference between the out-of-pair couplings:

$$\nu_1^{SLIC} = |2J_{intra}|$$
$$\tau_{SLIC} = 1/(\sqrt{2} \cdot \Delta J)$$
Eq. 2

The SLIC pulse is applied on resonance with the chemical shift of the $CH_2$-group of interest. Poly-SLIC can be used to irradiate several $CH_2$ groups in the same molecule simultaneously. The conditions of the SLIC pulse are different in this case, as they require a weaker amplitude of $\nu_1^{poly-SLIC} = J_{intra}$ and a longer duration of $\tau^{poly-SLIC} = 1/\Delta J$ [30].

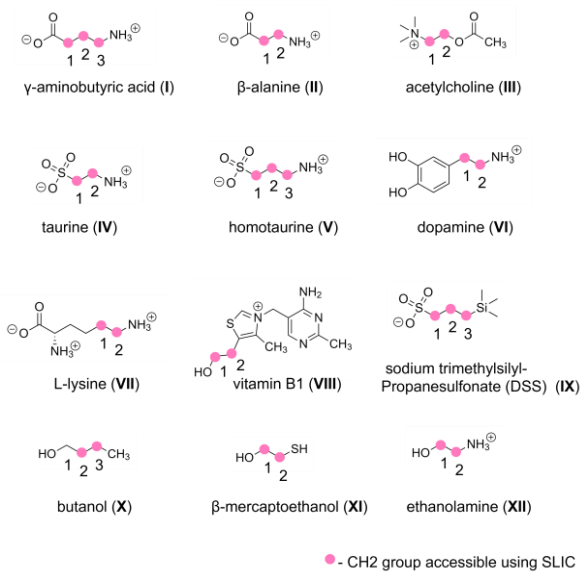

**Figure 1.** A selection of 12 common achiral and chiral molecules with up to four consecutive methylene groups. The CH$_2$ groups that are accessible for excitation of LLS by SLIC are coded in pink. In the case of (chiral) lysine (**VII**), the diastereotopic βCH$_2$ and γCH$_2$ groups feature distinct chemical shifts, so that LLS can also be excited by conventional methods, while the δCH$_2$ and εCH$_2$ groups have nearly degenerate chemical shifts, where LLS can be best excited by SLIC.

The parameters for SLIC pulses with a single irradiation frequency were optimized for all twelve molecules of **Figure 1**, as reported in **Table 1**. The optimum RF amplitudes $25 < \nu_1^{SLIC} < 29$ Hz are similar for all compounds, since the geminal *J*-couplings extend over a small range $12.5 < |^2J| < 14.5$ Hz. The optimum pulse durations vary over a broader range 110 ms $< \tau_{SLIC} <$ 844 ms. In the case of β-mercaptoethanol with a $\tau_{SLIC}$ of 844 ms, this reflects the small difference $\Delta J = 0.8$ Hz. In such cases, it is not immediately apparent from the appearance of the conventional spectrum that one is dealing with an $AA'XX'$ rather than an A$_2$X$_2$ system. The weaker the magnetic inequivalence, the more the multiplets take the appearance of first-order triplets with binomial 1:2:1 intensities. Despite their resemblance with weakly coupled systems, the CH$_2$ protons in β-mercaptoethanol, taurine and β-alanine can be addressed by SLIC.

*Rotational conformers and magnetic inequivalence*

In aliphatic chains comprising adjacent CH$_2$ units, the parameter $\Delta J$ (and hence the optimum duration $\tau_{SLIC}$) reflect the unequal populations of rotational conformers. These populations are affected by the following:

(i) Steric hindrance between bulky groups can cause one of the rotational conformers to be preferred. Compounds with less bulky substituents and shorter aliphatic chains, or weaker intramolecular interactions are therefore less likely to have accessible LLS. For example, no LLS could be excited by SLIC in propanol, but all CH2 groups in butanol **X** and pentanol except for the CH$_2$OH group have accessible LLS.

(ii) Electrostatic interactions between charged groups can stabilize one of the rotational conformers, thus granting access to LLS. At neutral pH, LLS can be excited in zwitterionic compounds like GABA, taurine and β-alanine. In these molecules, the LLS are no longer accessible when their amino groups are deprotonated at basic pH > 10.

(iii) If a compound contains H-bond donors and acceptors, some rotational conformers can be stabilized by intramolecular H-bonds.

**Table 1.** Optimized SLIC parameters and relaxation times T$_1$ and T$_{LLS}$ of various molecules.

| Molecule | See Fig. 1 | $\nu_1^{SLIC}$ [Hz] | $\tau_{SLIC}$ [ms] | Irradiated CH$_2$ (Fig.1) | $T_1$ [s] | $T_{LLS}$ [s] | $T_{LLS}/T_1$ |
|---|---|---|---|---|---|---|---|
| GABA | I | 27 | 210 | 1 | 2.2 | 11.2 | 5.1 |
|  |  |  |  | 2 | 2.5 | 10.3 | 4.1 |
| β-Alanine | II | 29 | 540 | 1 | 3.0 | 13.3 | 4.4 |
|  |  |  |  | 2 | 3.0 | 13.3 | 4.4 |
| Acetylcholine | III | 29 | 117 | 1 | 2.0 | 5.8 | 2.9 |
|  |  |  |  | 2 | 2.2 | 5.7 | 2.6 |
| Taurine | IV | 28 | 520 | 1 | 3.2 | 18.2 | 5.7 |
|  |  |  |  | 2 | 3.4 | 16.3 | 4.8 |
| Homotaurine | V | 26 | 180 | 1 | 1.9 | 10.2 | 5.4 |
|  |  |  |  | 2 | 1.9 | 9.7 | 5.1 |
|  |  |  |  | 3 | 2.0 | 8.2 | 4.1 |
| Dopamine | VI | 27 | 495 | 1 | 1.0 | 5.9 | 5.9 |
|  |  |  |  | 2 | 1.1 | 5.4 | 4.9 |
| L-Lysine | VII | 27 | 140 | 1 | 0.8 | 2.9 | 3.6 |
|  |  |  |  | 2 | 1.0 | 2.8 | 2.0 |
| Vitamin B1 | VIII | 26 | 250 | 1 | 0.9 | 6.0 | 6.7 |
|  |  |  |  | 2 | 0.9 | 4.5 | 5.0 |
| DSS | IX | 27 | 110 | 1 | 1.7 | 7.4 | 4.4 |
|  |  |  |  | 2 | 1.7 | 9.2 | 5.4 |
|  |  |  |  | 3 | 1.6 | 6.8 | 4.3 |
| Butanol | X | 25 | 195 | 1 | 0.9 | 6.0 | 6.7 |
|  |  |  |  | 2 | 0.9 | 4.5 | 5.0 |
| β-Mercaptoethanol | XI | 25 | 844 | 1 | 5.6 | 25.4 | 4.5 |
|  |  |  |  | 2 | 5.7 | 25.8 | 4.5 |
| Ethanolamine | XII | 26 | 365 | 1 | 3.6 | 9.0 | 2.5 |
|  |  |  |  | 2 | 3.6 | 8.4 | 2.3 |

*Polychromatic SLIC applied to mixtures*

The poly-SLIC approach can be applied to observe LLS simultaneously in different molecules in a mixture. **Figure 2a** shows a conventional NMR spectrum of a mixture containing DSS, ethanolamine, β-alanine, taurine and GABA. **Figure 2b** shows a spectrum of the same mixture after excitation of LLS and reconversion it into observable magnetization using a superposition of 5 poly-SLIC sequences with a common RF amplitude $\nu_1^{SLIC} = 27$ Hz and a compromise duration $\tau_{SLIC} = 319$ ms. Within each molecule, only one methylene group was irradiated. The integrated intensities of the signals varied between 0.5 and 1.2 % of their amplitudes in the conventional spectrum. For β-alanine, the signal is weak because $\nu_1^{SLIC}$ and $\tau_{SLIC}$ were rather far from their optimum values shown in T**able 1**.

In **Figure 2b**, only signals that stem from LLS are observed. All other peaks are removed from the spectrum, in particular strong peaks of water and methyl protons of DSS. Thus the proposed poly-SLIC sequence acts like a singlet-state filter[20,22]. The filter can be further adjusted to obtain high selectivity with respect to only one molecule of interest. This can be achieved by using poly-SLIC excitation of delocalized LLS in several $CH_2$ groups of one molecule. This filter relies on low probability of having several overlapping methylene signals in two different molecules.

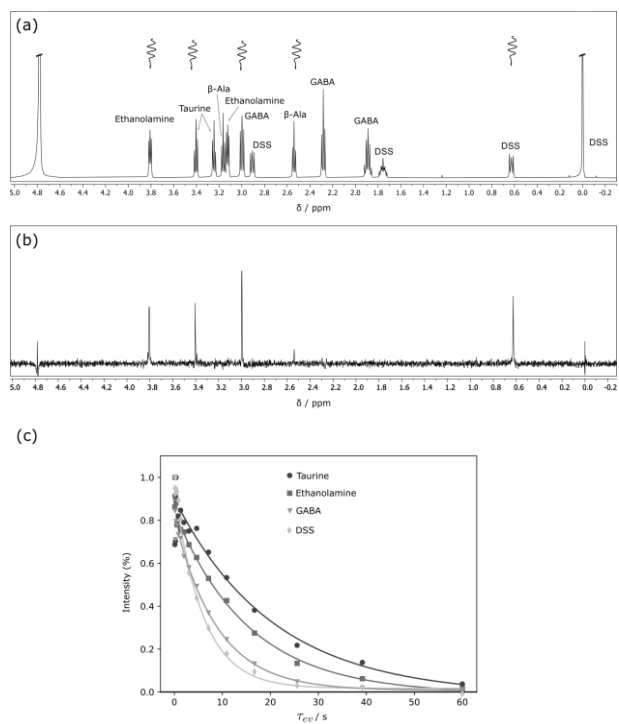

**Figure 2**: Simultaneous observation of LLS in a mixture containing *ca.* 10 mM each of compounds **I**, **II**, **IV**, and **XII** (**Figure 1** and **Table 1**.) (a) Conventional $^1$H NMR spectrum of the mixture acquired with 16 scans. Arrows specify the 5 frequencies at which selective SLIC pulses were applied. (b) A spectrum acquired after a poly-SLIC sequence with a LLS relaxation interval $\tau_{rel}$ = 3 s (16 scans). (c) LLS decays of four molecules measured simultaneously. The LLS lifetimes were determined from mono-exponential fits, ignoring weak initial oscillations that can be neglected after $\tau_{rel}$ > 1.3 s: $T_{LLS}$ (taurine) = 20.6 ± 2.3 s, $T_{LLS}$ (ethanolamine) = 15.4 ± 0.7 s, $T_{LLS}$ (GABA) = 8.9 ± 0.3 s, and $T_{LLS}$ (DSS) = 5.9 ± 0.5 s.

The decay curves of the four LLS signals were obtained simultaneously (**Figure 2c**) and their lifetimes $T_{LLS}$ were determined from mono-exponential fits. The observed lifetimes $T_{LLS}$ of the $CH_2$ groups in different molecules in **Figure 2c** range from *ca.* 6 s for DSS to 20 s for taurine. This can be partly explained by different values of $\Delta J$ in those molecules which cause singlet-triplet leakage. But comparison of $T_{LLS}$ measured at exactly same conditions and in the same type of group may also provide insight into the stochastic fluctuations that are responsible for relaxation of methylene groups. For example, the shortest-lived LLS in DSS stems from the $CH_2$ group which is located next to the $Si(CH_3)_3$ group (position 3 in **Figure 1**). There are nine methyl protons that can contribute to LLS relaxation through dipole-dipole interactions. Ethanolamine and taurine on the other hand have only two $CH_2$ groups, which are magnetically more silent and hence contribute less to LLS relaxation. This may provide information about (intra-) molecular mobility, and to the accessibility of paramagnetic oxygen and other radicals dissolved in the solvent. Rationalizing such differences in the remaining molecules shown in Figure 1 goes beyond the scope of this work.

## Conclusions

Long-lived states involving geminal pairs of protons in $CH_2$ groups in achiral aliphatic chains have been excited and observed by poly-SLIC. The lifetimes $T_{LLS}$ of the long-lived states are found to be significantly longer than $T_1$ and need not be sustained by any radio frequency fields at high magnetic fields. The discovery that LLS can be excited and observed in taurine, homotaurine, GABA, dopamine, and acetylcholine opens the opportunity of combining singlet state NMR methods with MRI in order to detect biologically active molecules and neurotransmitters in brain, which might broaden the scope of MRI[33]. While the quest for the *longest possible lifetimes* is a challenging objective, one may also seek the *highest possible contrast* between the lifetimes $T_{LLS}$ of a given molecule in different environments, such as a drug (or a competitor) that is either freely tumbling in solution, or partly bound to a macromolecular target (protein, nucleic acids), where $T_{LLS}^{bound} \ll T_{LLS}^{free}$, thus providing a strong contrast upon binding. LLS involving protons are particularly suitable for such purposes. This should improve titration experiments aimed at determining dissociation constants[32].

## Experimental Section

The spectra were obtained with a Bruker 500 MHz WB system equipped with 'Neo' transmitters that can generate pulse sequences such as shown in **Figure 3**, a 10 mm BBO probe, and a 5 mm iProbe. The $T_{00}$ filter allows one to eliminate undesirable peaks, such as the signals of HOD and of the methyl groups of DSS. The concentrations of the compounds in the mixture were about 10 mM. The solution was adjusted to pH 7 with a 50 mM phosphate buffer. Multiple frequencies for polychromatic SLIC were generated by phase modulation.

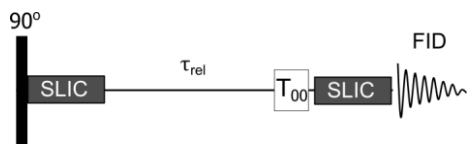

**Figure 3.** Pulse sequence to study relaxation of LLS of a chosen $CH_2$ group in chains of methylene groups. A 90° pulse is followed by SLIC that converts magnetization into the LLS. The decay of LLS is observed by acquiring a set of 1D spectra with a variable $\tau_{rel}$ interval; a $T_{00}$ filter; the second SLIC converts LLS back to magnetization which is then detected. Phase cycling of the SLIC pulses together with the receiver phase eliminates all signals that do not stem from LLS. In poly-SLIC of mixtures, several such sequences with different RF frequencies (obtained by phase modulation) are applied simultaneously to $CH_2$ groups of different molecules.

## Acknowledgements

We are indebted to Dr. Ulric le Paige, and Dr. Nicolas Birlirakis for stimulating discussions. We acknowledge the CNRS and ENS for support, and the European Research Council (ERC) for the Synergy grant "Highly Informative Drug Screening by Overcoming NMR Restrictions" (HISCORE, grant agreement number 951459).

**Keywords:** Long-lived states, Chemically equivalent protons, Spin-Lock Induced Crossing (SLIC), Aliphatic chains, Metabolites, Neurotransmitters